\documentclass[twocolumn,aps,superscriptaddress,prl]{revtex4-2}
\usepackage{graphicx}
\usepackage{xcolor}
\usepackage{times}
\usepackage{amsmath}
\usepackage{amssymb}
\usepackage{units}
\usepackage[normalem]{ulem}

\usepackage{natbib,hyperref}

\usepackage{float}
\usepackage{stmaryrd} 
\DeclareGraphicsExtensions{.pdf,.png,.eps,.jpg}

\begin{document}
\preprint{0}

\title{Observation of flat $\Gamma$ moir\'e bands in twisted bilayer WSe$_2$}

\author{G. Gatti} 
\thanks{These authors contributed equally to this work.}
\affiliation{Department of Quantum Matter Physics, University of Geneva, 24 Quai Ernest-Ansermet, 1211 Geneva, Switzerland} 

\author{J. Issing} 
\thanks{These authors contributed equally to this work.}
\affiliation{Department of Quantum Matter Physics, University of Geneva, 24 Quai Ernest-Ansermet, 1211 Geneva, Switzerland} 

\author{L. Rademaker} 
\affiliation{Department of Theoretical Physics, University of Geneva, 24 Quai Ernest-Ansermet, 1211 Geneva, Switzerland} 

\author{F. Margot} 
\affiliation{Department of Quantum Matter Physics, University of Geneva, 24 Quai Ernest-Ansermet, 1211 Geneva, Switzerland} 

\author{T. A. de Jong} 
\affiliation{Huygens-Kamerlingh Onnes Laboratory, Leiden Institute of Physics, Leiden University, Leiden, The Netherlands} 

\author{S. J. van der Molen } 
\affiliation{Huygens-Kamerlingh Onnes Laboratory, Leiden Institute of Physics, Leiden University, Leiden, The Netherlands} 

\author{J. Teyssier} 
\affiliation{Department of Quantum Matter Physics, University of Geneva, 24 Quai Ernest-Ansermet, 1211 Geneva, Switzerland} 

\author{T. K. Kim} 
\affiliation{Diamond Light Source, Harwell Campus, Didcot, OX11 0DE, United Kingdom} 

\author{M. D. Watson} 
\affiliation{Diamond Light Source, Harwell Campus, Didcot, OX11 0DE, United Kingdom} 

\author{C. Cacho} 
\affiliation{Diamond Light Source, Harwell Campus, Didcot, OX11 0DE, United Kingdom} 

\author{P. Dudin}
\affiliation{Synchrotron SOLEIL, L’Orme des Merisiers, Saint Aubin-BP 48, 91192 Gif sur Yvette Cedex, France}

\author{J. Avila}
\affiliation{Synchrotron SOLEIL, L’Orme des Merisiers, Saint Aubin-BP 48, 91192 Gif sur Yvette Cedex, France}

\author{K. Cordero Edwards} 
\affiliation{Department of Quantum Matter Physics, University of Geneva, 24 Quai Ernest-Ansermet, 1211 Geneva, Switzerland} 

\author{P. Paruch} 
\affiliation{Department of Quantum Matter Physics, University of Geneva, 24 Quai Ernest-Ansermet, 1211 Geneva, Switzerland} 

\author{N. Ubrig} 
\affiliation{Department of Quantum Matter Physics, University of Geneva, 24 Quai Ernest-Ansermet, 1211 Geneva, Switzerland}
\affiliation{Group of Applied Physics, University of Geneva, 24 Quai Ernest Ansermet, CH-1211 Geneva, Switzerland}

\author{I. Guti\'errez-Lezama} 
\affiliation{Department of Quantum Matter Physics, University of Geneva, 24 Quai Ernest-Ansermet, 1211 Geneva, Switzerland} 
\affiliation{Group of Applied Physics, University of Geneva, 24 Quai Ernest Ansermet, CH-1211 Geneva, Switzerland}

\author{A. Morpurgo} 
\affiliation{Department of Quantum Matter Physics, University of Geneva, 24 Quai Ernest-Ansermet, 1211 Geneva, Switzerland} 
\affiliation{Group of Applied Physics, University of Geneva, 24 Quai Ernest Ansermet, CH-1211 Geneva, Switzerland}

\author{A. Tamai} 
\affiliation{Department of Quantum Matter Physics, University of Geneva, 24 Quai Ernest-Ansermet, 1211 Geneva, Switzerland} 

\author{F. Baumberger}
\affiliation{Department of Quantum Matter Physics, University of Geneva, 24 Quai Ernest-Ansermet, 1211 Geneva, Switzerland} 

\date{\today}

\begin{abstract}
The recent observation of correlated phases in transition metal dichalcogenide moir\'e systems at integer and fractional filling promises new insight into metal-insulator transitions and the unusual states of matter that can emerge near such transitions.
Here, we combine real- and momentum-space mapping techniques to study moir\'e superlattice effects in 57.4$^{\circ}$ twisted WSe$_2$ (tWSe$_2$). Our data reveal a split-off flat band that derives from the monolayer $\Gamma$ states. Using advanced data analysis, we directly quantify the moir\'e potential from our data.
We further demonstrate that the global valence band maximum in tWSe$_2$ is close in energy to this flat band but derives from the monolayer K-states which show weaker superlattice effects.
These results constrain theoretical models and open the perspective that $\Gamma$-valley flat bands might be involved in the correlated physics of twisted WSe$_2$.
\end{abstract}

\maketitle

The bonding characteristics of van der Waals materials naturally lead to moir\'e superlattices at interfaces with a lattice mismatch or small twist angle. Exploiting this tuning knob in graphene, WSe$_2$ and other transition metal dichalcogenides (TMDs) revealed a wealth of non-trivial many-body phases that are not observed in the parent materials. Examples include Mott- and Mott-Wigner like insulating states~\cite{Regan2020,Tang2020,Xu2020,Wang2020,Ghiotto2021,Li_Mak2021,Shimazaki2020,Li_Crommie2021}, exciton insulators~\cite{Xu2022,Zhang2022}, magnetism~\cite{Sharpe2019,Xu2022}, superconductivity~\cite{Cao2018,Lu2019} and strange metal phases~\cite{Ghiotto2021,Jaoui2022}.

The correlated phases in semiconductor moir\'e systems
are generally attributed to strong correlations in monolayer K-states that form flat minibands in the moir\'e superlattice. However, imaging such mini bands in momentum space proved challenging and key-questions such as the band width, the magnitude of the mini gaps and even from which monolayer valley the mini bands derive remain largely open.

\textit{Ab-initio} calculations of tWSe$_2$ and other twisted TMDs are thus far limited to isolated bilayers without substrate or encapsulation and do not include doping or displacement fields~\cite{Naik2018,Angeli2021,Vitale2021,Zhang2021,Devakul2021,Kundu2022}. Moreover, they disagree over the strength of superlattice effects and 
the nature of the mini-bands at the top of the valence band.
Most natural 2H TMD bilayers have the valence band maximum (VBM) at the $\Gamma$-point where states derive from the antibonding metal $d_{z^2}$ / chalcogen $p_{z}$ orbital. However, in 2H WSe$_2$ (and MoTe$_2$) the large spin-orbit splitting of the metal $d_{xy}$ / $d_{x^2-y^2}$ derived states at K shifts the global VBM to the K-point~\cite{Wilson2017,Lindlau2018}. Recent large scale density functional calculations find that twist angles below $\approx 3^{\circ}$ tip this balance and move the VBM of tWSe$_2$ back to $\Gamma$~\cite{Vitale2021}. 
A VBM at $\Gamma$ was also reported in an ARPES study of twisted double bilayer WSe$_2$~\cite{An2020}.
Other calculations for different twist angles~\cite{Angeli2021,Devakul2021,Kundu2022} and a recent experiment at $5.1^{\circ}$ twist angle~\cite{Pei2022}, find that the VBM derives from monolayer K states. 

The nature of the VBM states is crucial because the effective low-energy models used to study correlation effects in twisted TMDs differ for $\Gamma$ and K states.
For $\Gamma$ derived states, theoretical studies find an emergent honeycomb symmetry with a moir\'e mini-band structure that resembles artificial graphene~\cite{Vitale2021,Angeli2021,Devakul2021,Kundu2022}. For K-derived states, on the other hand, recent theoretical work found a parabolic moir\'e VBM at K on the mini-Brillouin zone~\cite{Vitale2021} and an electric field induced topological insulator phase~\cite{Wu2019}. 
These differences have obvious implications for the interpretation of transport measurements that are currently being performed and that exhibit a plethora of exciting new phenomena.

The strength of superlattice effects is commonly parameterized by an effective moir\'e potential.
However, estimates of the moir\'e potential in the literature show a considerable spread ranging from a few milli electron volt (meV) up to hundreds of meV~\cite{Angeli2021,Shabani2021,Geng2020,Zhang2021}.
Previous $\mu$-ARPES studies on tWSe$_2$ and other TMD moir\'e systems found signatures of replica bands shifted by the moir\'e reciprocal lattice vector but could not resolve flat moir\'e mini bands 
or estimate the effective moir\'e potential~\cite{Ulstrup2020,Xie2020,Pei2022}.
Scanning tunneling microscopy (STM) studies of tWSe$_2$ found that the local density of states is spatially modulated and, depending on energy and twist-angle, can have triangular, honeycomb or Kagome symmetry~\cite{Zhang2020,Li2021,Pei2022}. STM further revealed a rich structure of tunneling spectra in the valence band consistent with the formation of moir\'e minibands. Correlating the real space information from STM with band structure calculations is, however, non-trivial.

\begin{figure*}[t!]
  \centering   \includegraphics[width=0.97\textwidth]{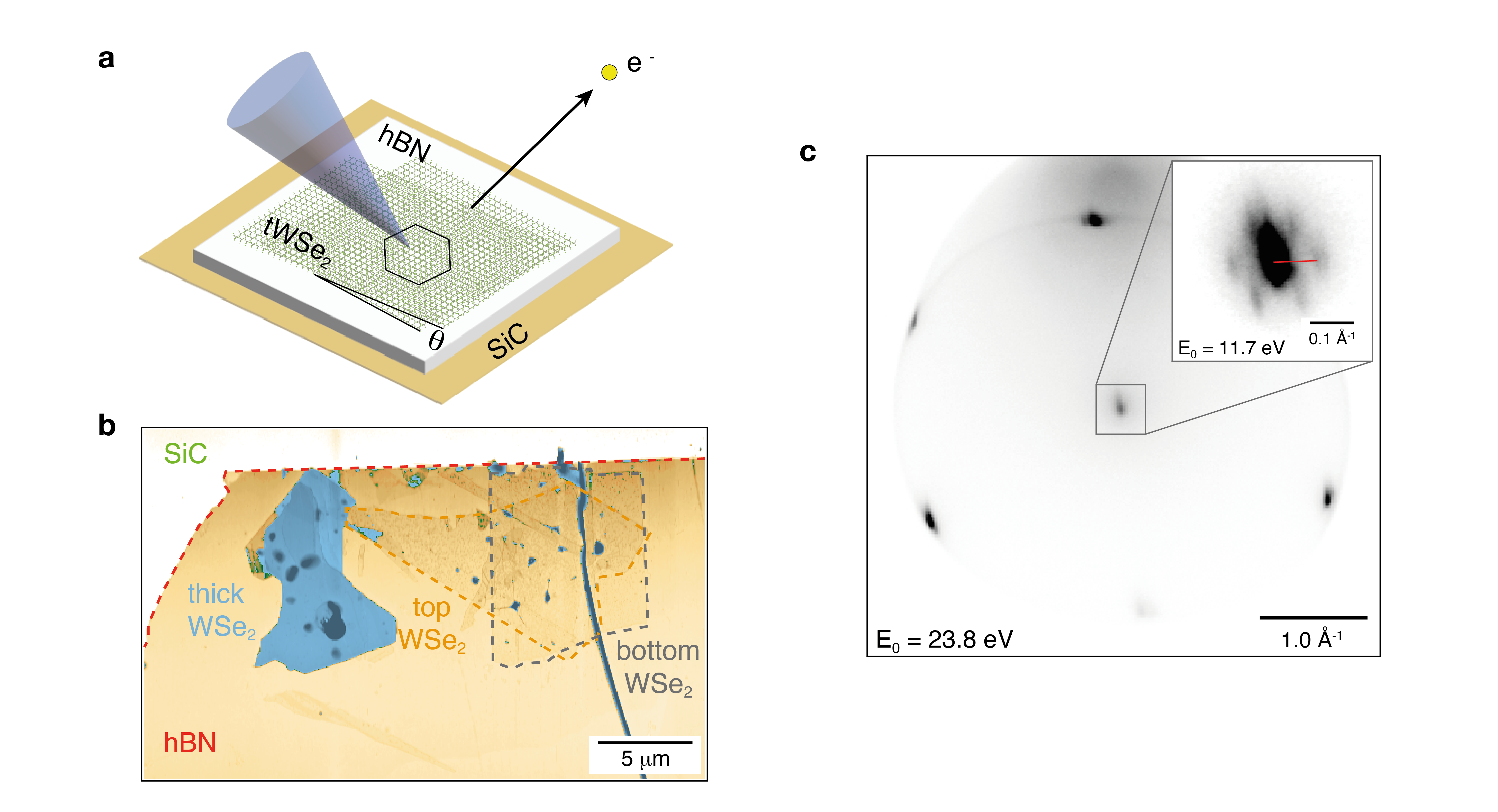}
  \caption[didascalia]{{\bf Device layout and characterization} (a) Schematic of the $\mu$-ARPES experiment on tWSe$_2$. Heterostructures of tWSe$_2$ on hBN were produced with a tear and stack method and studied with synchrotron based micro-focus ARPES (for details, see Methods). The black hexagon indicates the moir\'e unit cell for an arbitrary twist angle $\theta$. (b) Topographic atomic force microscopy image of the van der Waals heterostructure. The scale bar is 5~$\mu$m. (c) $\mu$-LEED pattern of the twisted bilayer WSe$_2$. The inset shows a zoom of the (0,0) spot with the satellite peaks used to quantify the moir\'e wavelength. The red line indicates a reciprocal moir\'e superlattice vector.
 }
  \label{fig1}
\end{figure*}

Here, we report the direct observation of a flat moir\'e mini band in tWSe$_2$. The flat band derives from the monolayer $\Gamma$-states and is in close proximity to the global VBM opening the possibility that the $\Gamma$-valley flat band participates in the correlated phases of tWSe$_2$. We estimate the band width of the moir\'e mini-band and determine the moir\'e potential from a quantitative analysis of the ARPES data.

\begin{figure*}[t!]
    \centering \includegraphics[width=0.98\textwidth]{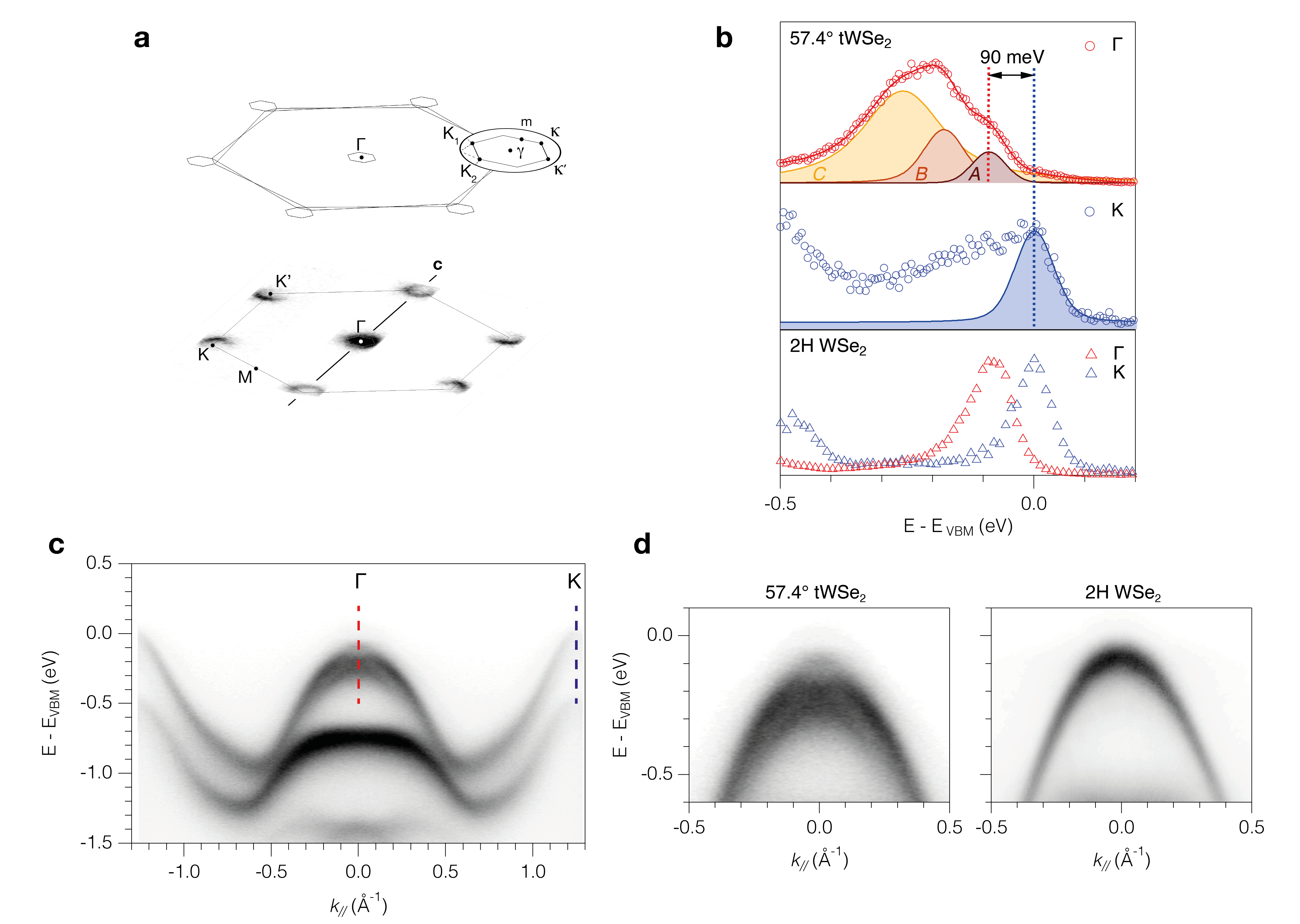}
  \caption[didascalia]{{\bf Valence band maximum in tWSe$_2$.} (a) ARPES constant energy map 200~meV below the VBM. The Brillouin zones (BZs) of the two twisted monolayers are indicated by grey hexagons. Small hexagons show the moir\'e mini BZs. (b) Energy distribution curves (EDCs) from tWSe$_2$ and untwisted 2H WSe$_2$ extracted at $\Gamma$ and K of the monolayer BZ. The EDCs from tWSe$_2$ are fit with three and one Lorentzians, respectively, convolved with a Gaussian of 70~meV full width half maximum to account for energy resolution and inhomogeneous broadening.
  (c) ARPES band dispersion of tWSe$_2$ measured along $\Gamma$K over the large monolayer BZ. (d) Zoom-in of the ARPES spectral functions of twisted and untwisted WSe$_2$ near the $\Gamma$ point. 
    }
  \label{fig2}
\end{figure*}

Fig.~1 illustrates the methodology used in our study. We fabricate the tWSe$_2$ structure with the tear and stack method using a polypropylene carbonate (PPC) stamp, first picking up an hBN flake and then the WSe$_2$ layers~\cite{Kim2016}. Prior to the ARPES measurements, we invert the stack, release it on a graphitized SiC substrate and sublimate the PPC in vacuum. 
This technique avoids any contact of tWSe$_2$ with polymers, which facilitates obtaining a clean, polymer-free sample surface.
We determine the precise twist angle of the heterostructure used for the $\mu$-ARPES experiments from micro-focus low-energy electron diffraction ($\mu$LEED) (see Fig.~1).
We first use the three-fold symmetry of the $\mu$LEED patterns in TMDs~\cite{jong_measuring_2018} to independently confirm the stacking angle near $\approx 60^{\circ}$, corresponding to a 2H structure. 
A zoom-in around the $(0,0)$ specular beam reveals six satellite spots that define a moir\'e mini Brillouin zone (BZ) with a reciprocal lattice vector $\mathrm{G}^{\mathrm{Moir\acute{e}}}\approx 1$~nm$^{-1}$ corresponding to a moir\'e wavelength of $a_M = 7.3$~nm and a twist angle $\theta = 57.4\,\pm\,0.1^{\circ}$. Note that the moir\'e mini-BZ is rotated by $90^{\circ}$ with respect to the atomic BZ defined by the main diffraction spots. 
Measurements at multiple locations on the tWSe$_2$ indicate a twist angle variation of no more than 0.4$^{\circ}$ in the sample studied here.

Figure \ref{fig2} shows the valence band dispersion of tWSe$_2$ over the entire monolayer Brillouin zone. 
At first glance, the overall dispersion is reminiscent of a natural 2H bilayer with parabolic band maxima at $\Gamma$ and K~\cite{Wilson2017}. However, a closer inspection of the data near $\Gamma$ (Fig.~\ref{fig2}d) reveals the presence of additional features on a much smaller energy scale that are absent in untwisted 2H WSe$_2$.
Extracting energy distribution curves (EDC) at $\Gamma$ from the data in Fig.~2d, we resolve three distinct components in tWSe$_2$ (features \textit{A, B, C} in Fig.~2b), in stark contrast to the EDC of 2H WSe$_2$, which shows a single peak only.
This splitting into multiple peaks in tWSe$_2$ is a fingerprint of moir\'e mini bands.
It also shows that moir\'e superlattice effects alter valley energies.
None of the 3 components in tWSe$_2$ can be trivially associated with the single peak in 2H WSe$_2$. This suggests that moir\'e effects can shift valleys by several ten meV, which is non-negligible in bilayer WSe$_2$ because of the close proximity in energy of states at $\Gamma$ and K~\cite{Wilson2017,Lindlau2018,Nguyen2019}.

The spectrum at K shows a well defined onset but no clear multi-peak structure.
This implies weaker superlattice effects than at $\Gamma$, in qualitative agreement with theoretical predictions~\cite{Vitale2021,Angeli2021}.
We quantify valley energies from experiment by comparing the highest-energy feature $A$ in the multi-peak fit at $\Gamma$ with a fit of the onset of spectral weight at K.
This shows that the $\Gamma$-valley in tWSe$_2$ lies $90 \pm 20$~meV lower than the global VBM measured at K. This is comparable to the variation of valley energies with strain~\cite{Amin2014}, doping and displacement fields~\cite{Brumme2015} and only slightly larger than the chemical potential of WSe$_2$ at relevant densities, all suggesting the $\Gamma$-valley flat band might contribute to the correlated phases observed in tWSe$_2$~\cite{Wang2020,Ghiotto2021,Xu2022}.

\begin{figure*}[t!]
  \centering   \includegraphics[width=0.98\textwidth]{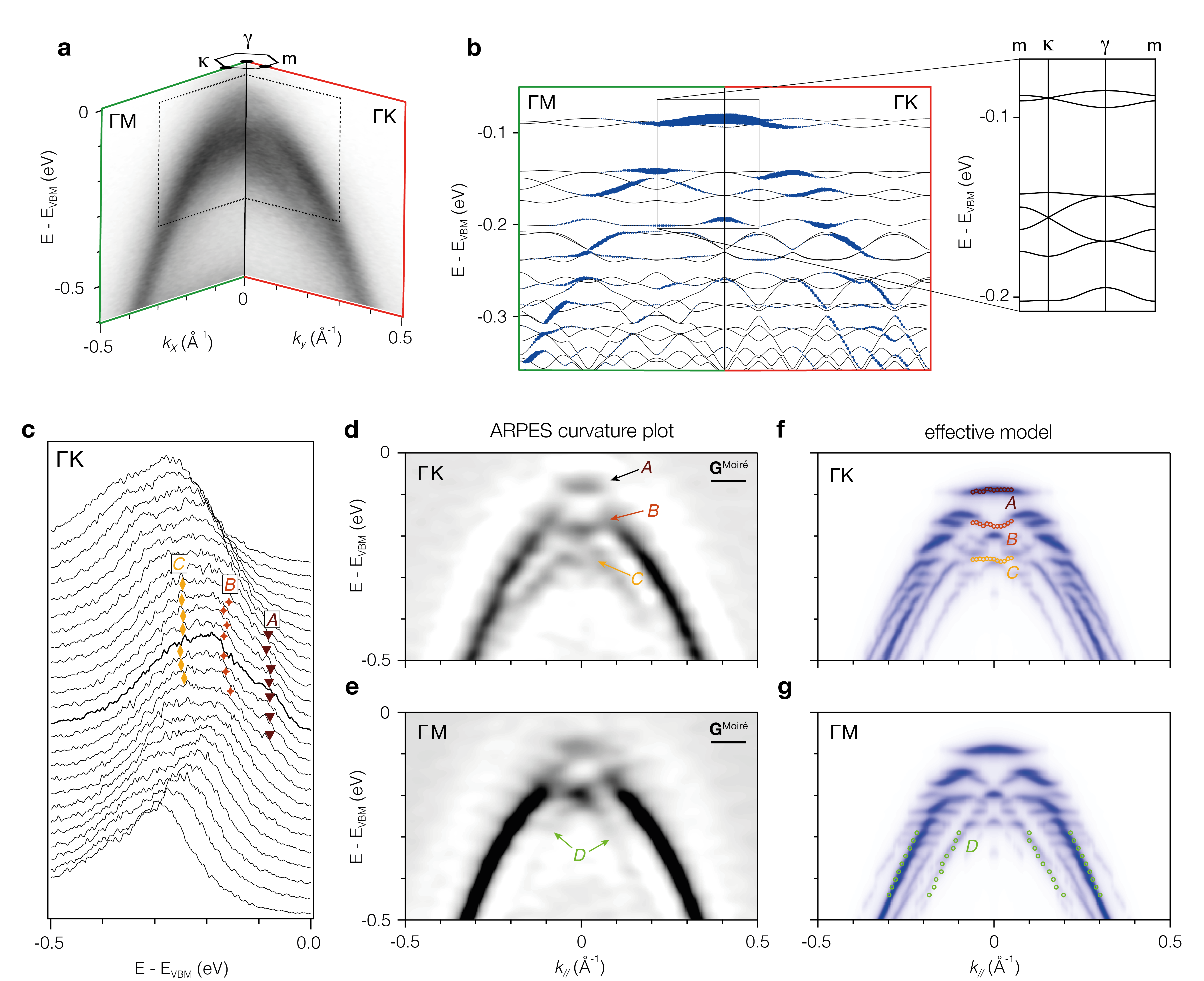}
  \caption[didascalia]{{\bf Moir\'e mini bands.} (a) ARPES band dispersions near the $\Gamma$ point along the two high symmetry directions. The small hexagon indicates the mini BZ.
  (b) Band dispersion (black) and spectral weights (blue) of the effective Hamiltonian in Eq.~\ref{H_eff} with parameters $V_0 = 48$~meV and $\phi = 121{^\circ}$.
  (c) Stack of EDCs in a momentum window of 0.2~\AA$^{-1}$ along K-$\Gamma$-K'. Markers indicate the position of the three Lorentzian peaks obtained from curve fits.
 (d,e) Curvature plots of the ARPES band dispersions along the K-$\Gamma$-K' and M-$\Gamma$-M directions, respectively. The scale bar corresponds to a moir\'e reciprocal lattice vector. 
 (f,g) Simulated spectral function of the effective Hamiltonian with the same parameters used in (c). For a visual comparison with the curvature plots in (d,e) we use a Lorentzian broadening of 10~meV which is smaller than the optimal value 
 found in the numerical optimization (see supplementary information, section~II).}
  \label{fig3}
\end{figure*}

Fig.~3a shows the complex spectral function of the moir\'e states along both high symmetry directions.
Near the $\Gamma$ point we can directly trace the dispersion of features \textit{A, B, C} from curve fits of the EDCs shown in Fig.~3c. Additional, more dispersive features at higher energy are evident in the raw image and curvature plots of the ARPES data.
The topmost feature $A$ defines the first moir\'e miniband. Consistent with the curvature plots, our fits show that this band is nearly dispersionless and separated by a gap of several 10~meV from other states. Tracing feature $A$ over the full mini BZ, we estimate a band width of the first moir\'e mini band of $\approx 10$~meV. This is small compared to the Hubbard interaction estimated as $U\approx e^2/(4\pi\epsilon\epsilon_0 d)\approx 80$~meV (assuming $\epsilon\approx 5$ and $d\approx a_M /2 = 3.6$~nm) placing half-filled $57.4^{\circ}$ tWSe$_2$ in the strongly correlated regime.

We now compare our data to a generic effective continuum model describing the motion of electrons in the bonding and antibonding valence band of twisted TMDs in the presence of an effective moir\'e potential~\cite{Zhang2021}:

\begin{eqnarray}
H & = &\left[ 
\begin{array}{cc}
-\frac{\hbar^2\mathbf{k}^2}{2m^{\star}} + V_1 & t_{\perp} (\mathbf{k}) \\  
t_{\perp} (\mathbf{k})  & \frac{\hbar^2\mathbf{k}^2}{2m^{\star}} + V_2 
\end{array}
\right] \nonumber \\
& & V_{1,2}  =  2V_0 \sum^3_{j=1} \cos(\mathbf{G}_j^{\mathrm{Moir\acute{e}}}\cdot \mathbf{r} \pm \phi)
\label{H_eff}
\end{eqnarray}

Here, the effective mass $m^{*}$ and interlayer hopping $t_{\perp}(\mathbf{k})$ parametrize the band structure of untwisted bilayers. 
We determine both of these parameters from fits to density functional theory calculations. Supplementary Fig.~5 shows that this is in excellent agreement with ARPES data from exfoliated 2H WSe$_2$.
The moir\'e physics is described with the effective potential $V_{1,2}$, which we expand in a first order Fourier series in the moir\'e wave vectors $\mathbf{G}_j^{\mathrm{Moir\acute{e}}}$ with phase factors $\phi$ that change sign with layer to be consistent with the 2-fold rotational symmetry with layer exchange.

This model reproduces the band structure found in previous theoretical work~\cite{Angeli2021,Zhang2021} with the characteristic Dirac cones at $\kappa$ in the first moir\'e mini-band for a wide range of phases $\phi$ (Fig.~3b). 

For a direct comparison with the ARPES data, we first calculate spectral weights by projecting the eigenstates of $H$ on plane wave states with momentum $\mathbf{k}$ (Fig.~3b,f,g)~\cite{Voit2000,Moser:2017es,Lisi2021}.
Performing calculations for different combinations of $V_0$, $\phi$ we identify three principal features that are sensitive to the model parameters: $i)$ As expected from theory, the gap between the first two moir\'e subbands is controlled by $V_0$; \textit{ii)} the spectral weight distribution of the dispersive features at high binding energy is very sensitive to the phase $\phi$, while \textit{iii)} the separation of features $A,B,C$ is determined by a combination of $V_0$ and $\phi$. 
Hence, $V_0$ and $\phi$ are sufficiently independent to be determined from our data. To do so in an unbiased way we calculate $\approx 5'000$ spectral functions for different parameters and determine the least square difference to the raw data. This yields $V_0=48 \pm 8$~meV and $\phi=121^{\circ} \pm 7^{\circ}$ where the error bars are estimated from the standard deviation of the parameters of all calculations weighted by the inverse of their square difference to the data.

Fig.~3f,g shows the continuum model spectral functions with the above determined parameters and a reduced broadening, demonstrating a good overall agreement with the data.
This confirms the interpretation of the main spectral features in terms of moir\'e mini-bands. The calculations clearly reproduce a split-off flat band that is highly localized in momentum space (feature $A$) suggesting a wave function extending over much of the moir\'e unit cell. 
The band width ($\Gamma$ to M dispersion) of the first subband in our model is $6$~meV, in fair agreement with our experimental estimate of 10~meV.
The calculations also reproduce the weakly dispersive features $B,C$ as well as the side band $D$. We note that the latter has been reported previously in Ref.~\cite{Pei2022} where its separation from the main band was identified with the moir\'e reciprocal lattice vector. Our calculations show that such an interpretation is not evident as a multitude of dispersive side bands are expected at various momenta.

The phase $\phi\approx120^{\circ}$ determined from our numerical analysis confirms the emergent honeycomb symmetry of the first subband, predicted by large scale density functional theory calculations~
\cite{Angeli2021,Zhang2021,Vitale2021}.
We note that a honeycomb charge distribution was also reported in some STM spectroscopic imaging experiments~\cite{Zhang2020,Pei2022} while another recent STM study found a triangular symmetry for the highest visible subband~\cite{Li2021}.

Our quantification of superlattice effects in tWSe$_2$ provides a benchmark for large scale DFT calculations.
More broadly, our work lends strong support to an interpretation of the remarkable transport and optical properties of tWSe$_2$ in terms of a correlated flat band system. However, our work also shows that tWSe$_2$ lacks the clear separation of energy scales needed for a description by a single band Hubbard model. The moir\'e potential of $V_0 = 48$~meV determined from our quantitative analysis is comparable to the energy difference of $\Gamma$ and K-valleys and also to shifts in valley energies predicted under realistic doping, displacement field and strain values~\cite{Brumme2015,Amin2014}. Moreover, for the twist angle studied here, the onsite interaction $U$ likely exceeds the moir\'e potential and thus the gap between subband manifolds. Accounting for this complexity will be essential for the interpretation of magneto-transport and optical measurements and will require the development of multi-band theoretical models.

\section*{ACKNOWLEDGMENTS}
This work was supported by the Swiss National Science Foundation (SNSF) under grants 184998, 178891. LR acknowledges support from a SNSF Ambizione fellowship 174208. TAdJ acknowledges financial support by the Netherlands Organisation for Scientific Research (NWO/OCW) as part of the Frontiers of Nanoscience (NanoFront) program.
We acknowledge Diamond Light Source for time on Beamline I05 under Proposal SI29021.

\section*{METHODS}

Commercially available hexagonal Boron Nitride (hBN) and WSe$_2$ have been exfoliated in air on 285~nm SiO$_2$/Si wafers. Crystal flakes with a specific thickness were recognized from their optical contrast under a microscope, and the thickness was later confirmed by AFM measurements. Heterostructures were assembled in a glove box using a motorized transfer system.
Transfer stamps were fabricated by placing a spin-coated PPC film on a dome-shaped handle made of polydimethylsiloxane (PDMS). 
To fabricate the heterostructure used in this work, we first picked up a large 17~nm thick hexagonal boron nitride (hBN) flake, whose sharp and straight edge was later employed to tear a monolayer WSe$_2$ flake in two parts while simultaneously picking up one of them. Subsequently, the stage was rotated by 57$^{\circ}$ and the remaining WSe$_2$ monolayer was picked up. Finally, the PPC film was detached from the PDMS base, flipped and released on a graphetized SiC substrate. Polymer residues and other contaminants were removed by annealing in ultra-high vacuum at 400$^{\circ}$C for 2 hours.

Photoemission experiments were performed at the nanoARPES endstation of the I05 beamline at Diamond Light Source, Didcot (UK) and the Antares beamline of Soleil synchrotron, France. The data shown in this manuscript were acquired at Diamond I05 using
80~eV light focused to $\approx 1\times1$~$\mu$m$^2$ spot size using a Fresnel zone plate. Immediately prior to the experiments, the sample was annealed in ultra-high vacuum at 400$^\circ$C for 30 minutes before transferring it \textit{in-situ} to the experimental chamber. The photoemission experiments were performed at 180~K using an energy and momentum resolution of $\approx 20$~meV and $\approx 0.01$~\AA$^{-1}$. 
Note that the tWSe$_2$ was not in direct electrical contact with the chamber ground. Nevertheless, charging effects remained well below the resolution of this experiment, which we attribute to the high photo-conductivity of hBN.

Low-Energy Electron Microscopy (LEEM) and $\mu$LEED were performed after photoemission experiments in the SPECS P90 based ESCHER system at Leiden University. Before measurements, the sample was reheated to $350^\circ$C (as measured by a pyrometer with emissivity set to 1.0) in the ultrahigh-vacuum of the measurement chamber (base pressure $<1.0\times 10^{-9}$mbar).
The sample was located on the substrate using photo-emission electron microscopy with an unfiltered mercury short-arc lamp. Initial $\mu$LEED measurements confirming the 2H-like geometry where performed using a $1.1\:\mu$m effective diameter circular illumination aperture. 
Subsequently the satellite diffraction spots arising from the moir\'e lattice were measured using a $0.23\:\mu$m effective diameter illumination aperture. Twist angle variation was determined by scanning the sample stage over the area used for $\mu$-ARPES experiments while recording diffraction patterns using the same illumination aperture.


\bibliography{twse2bibnew.bib}

\end{document}